\documentclass[3p,twocolumn]{elsarticle}

\usepackage{ecrc}

\volume{00}

\firstpage{1}

\journalname{Nuclear and Particle Physics Proceedings}

\runauth{G. Giacinti \& A. M. Taylor}

\jid{nppp}

\jnltitlelogo{Nuclear and Particle Physics Proceedings}

\usepackage{amssymb}

\usepackage{rotating}

\newcommand{\bfm}[1]{\mbox{\boldmath$ #1 $}}

\def\lsim{\mathrel{\rlap{\lower4pt\hbox{\hskip1pt$\sim$}}\raise1pt\hbox{$<$}}}

\begin{document}

\begin{frontmatter}



\dochead{}

\title{Galactic Cosmic-Rays in a Breeze}

\author[label1]{Gwenael~Giacinti}
\author[label2]{Andrew~M.~Taylor}

\address[label1]{Max-Planck-Institut f\"ur Kernphysik, Postfach 103980, 69029 Heidelberg, Germany}
\address[label2]{Dublin Institute for Advanced Studies, 31 Fitzwilliam Place, Dublin 2, Ireland}

\begin{abstract}
We study a scenario in which the Fermi bubbles are formed through a Galactocentric outflow of gas and pre-accelerated cosmic-rays (CR). We take into account CR energy losses due to proton-proton interactions with the gas present in the bubbles, and calculate the associated gamma-ray emission. We find that CRs diffusing and advecting within a breeze outflow result in an approximately flat surface brightness profile of the gamma-ray emission, as observed by Fermi satellite. Finally, we apply similar outflow profiles to larger Galactocentric radii, and investigate their effects on the CR spectrum and boron-to-carbon ratio. Hardenings can appear in the spectrum, even in cases with equal CR diffusion coefficients in the disk and halo~\cite{Taylor:2016qel}.
\end{abstract}


\end{frontmatter}


\section{Introduction}
\label{intro}


A number of indications that the center of our Galaxy feeds a wind has been found over the last few decades. This body of evidence has been provided from observations in a broad energy range: radio HI~\cite{Lockman:1984}, infrared (IR)~\cite{Morris:1996}, and X-rays~\cite{Cheng:1996}. IR observations have also indicated that this wind continues further away~\cite{Bland-Hawthorn:2003}, and that it may be responsible for the larger structures observed out of the Galactic plane. Absorption line features in the spectra of Active Galactic Nuclei (AGN) can be used as a probe of the structure of the gas flow: See Reference~\cite{Keeney:2006au}, whose results indicate the presence of a coherent gas flow, consistent with an outflow being directed away from the Galactic plane.

More recent gamma-ray and radio observations have shown the presence of extended non-thermal particle populations in bubble-like structures in the halo, both above and below the Galactic center (GC), see References~\cite{Su:2010qj,Yang:2014pia,Fermi-LAT:2014sfa,Carretti:2013sc}. The current picture seems to indicate that cosmic-rays (CR) and hot gas are conveyed out from the GC region into the halo within a Galactocentric outflow.

As for the velocity of this outflow, values of $\sim 300$\,km/s have been suggested in the region close to the Galactic disk (within $\sim$ a couple of kiloparsecs), from the weakness of the X-ray features associated with the edge of the bubbles~\cite{Su:2010qj,Kataoka:2013tma,Fang:2014hea,Fox:2015}. At distances $\sim 4$\,kpc and $\sim 9$\,kpc away of the Galactic plane, observations of high velocity clouds suggest velocities of about $\sim 150$\,km/s, cf. Reference~\cite{Keeney:2006au}. Further out, towards the edges, velocities are $\lesssim 100$\,km/s. In radio~\cite{Carretti:2013sc}, the bending observed in the outflow at high latitudes may be related to the motion of our Galaxy towards Andromeda (relative velocity $\sim 50$\,km/s).

In the present work, we study the secondary signatures produced by CRs embedded in outflows.

In Section~\ref{GC_outflow}, we focus on the Fermi bubbles. We then apply, in Section~\ref{CR_at_Earth}, a similar outflow velocity profile at larger Galactocentric radii, and investigate the possible traces it would leave on local CR observables, should such an outflow exist locally. We present our conclusions in Section~\ref{conclusion}.

\section{Fermi bubbles: CRs and gamma-rays associated with a Galactocentric Outflow}
\label{GC_outflow}

\begin{figure*}[t]
\includegraphics[width=0.45\linewidth]{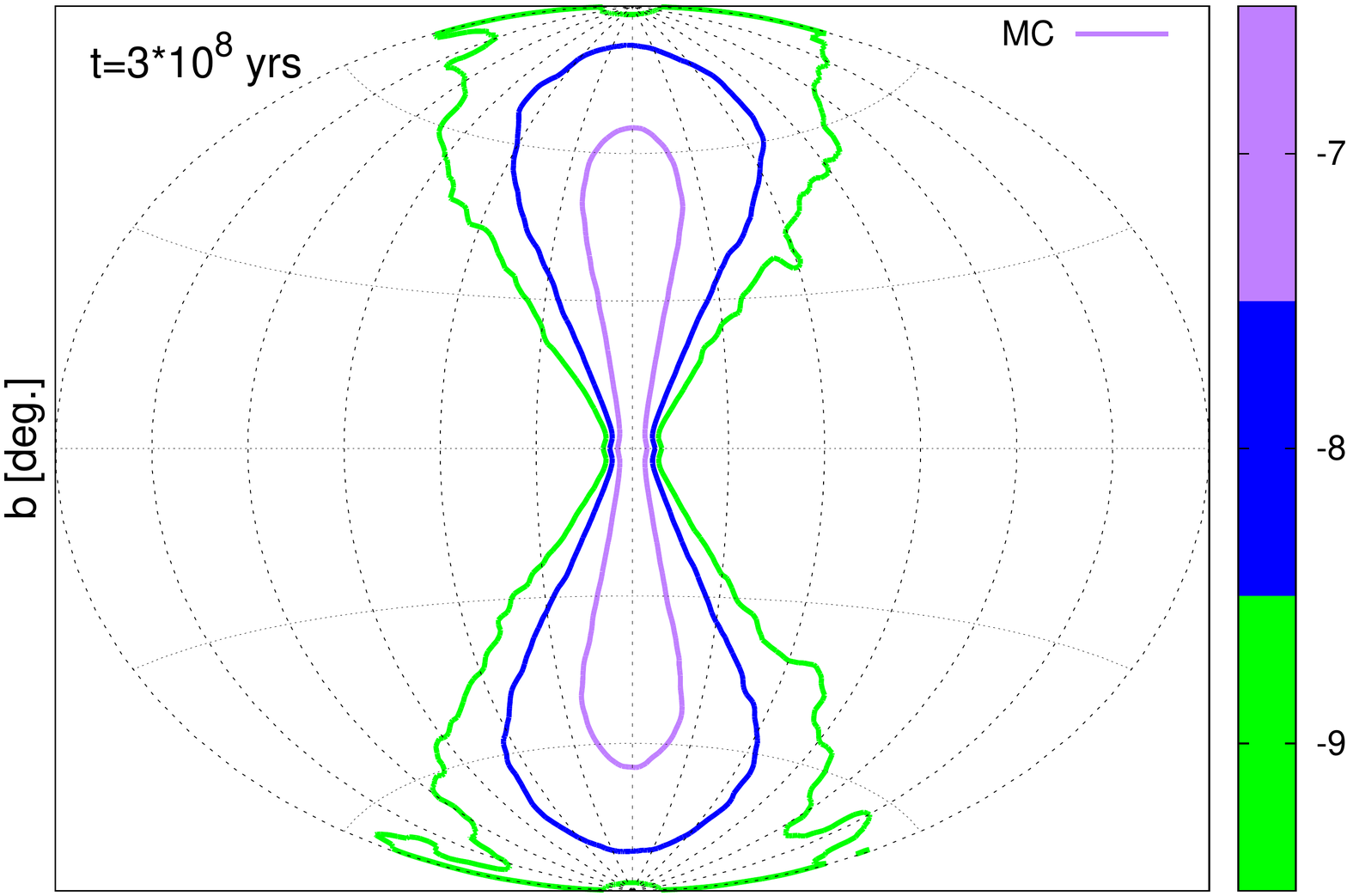}
\includegraphics[width=0.45\linewidth]{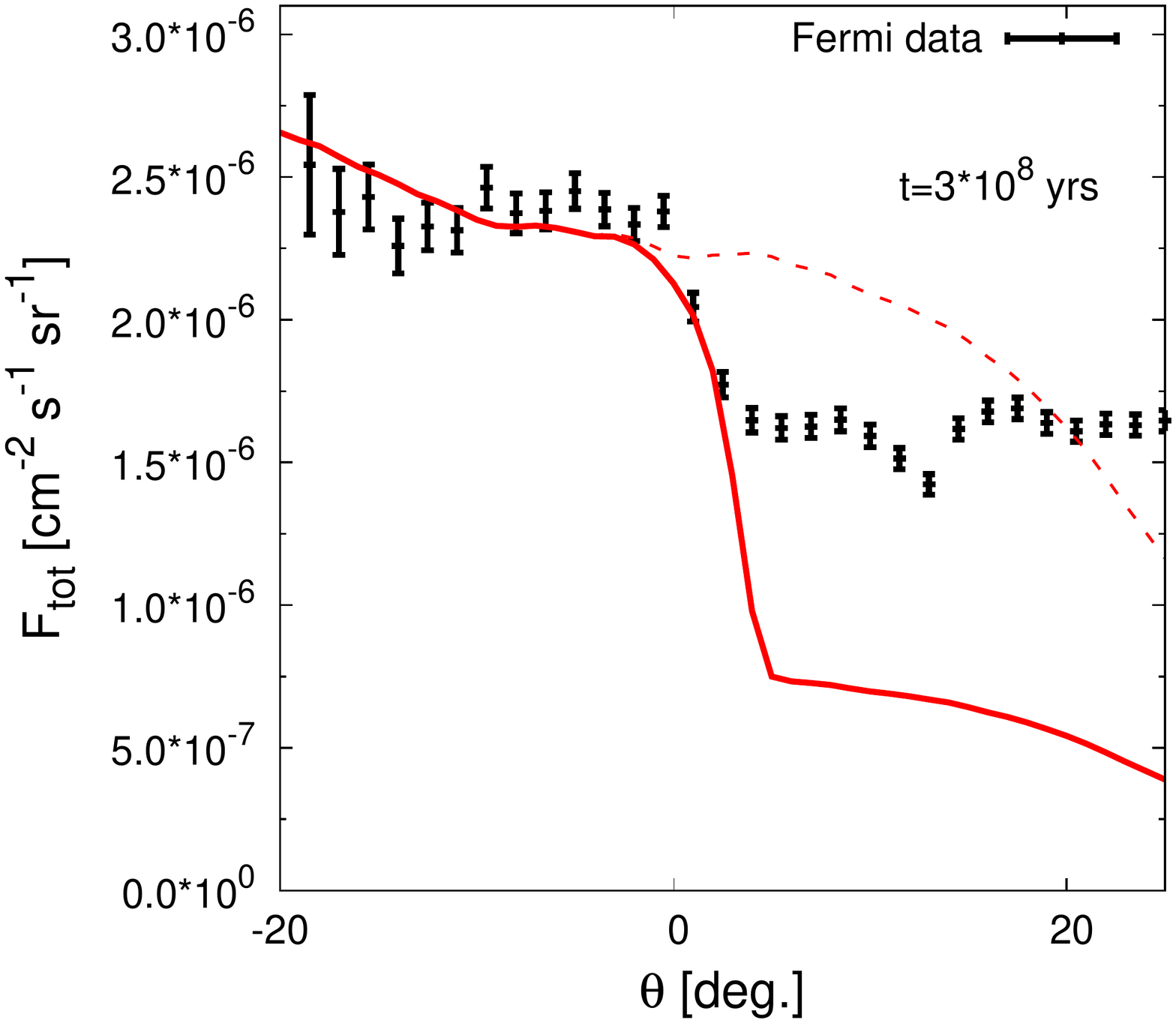}
\caption{{\it Left panel:} Contours showing the $\log_{10}$ of the gamma-ray flux surface brightness (in cm$^{-2}$\,s$^{-1}$\,sr$^{-1}$) from the bubbles, following CR interaction with the gas in the outflow. The different line colours indicate the corresponding contour values, cf. values in the colour bar. {\it Right panel:} Comparison of the edge of the $(1-2)$\,GeV gamma-ray bubble from our model with that from the Fermi observation analysis. We count the angle $\theta$ from the edge of the bubble. At large $\theta$, for the energy bin considered, further diffuse gamma-ray background~\cite{Abdo:2010nz} dominates the observed flux ---the model values sit below this level in this region. Solid line for a decrease in the gas density at the bubble edge, and dashed line for a constant density throughout.}
\label{gamma_contours}
\end{figure*}

A description for the propagation of cosmic rays in a turbulent
region in which an advective flow is present is provided by,
\begin{displaymath}
  \frac{\partial \psi_{\rm CR}}{\partial t} = \bfm{\nabla}\cdot \left( \mathcal{D} \bfm{\nabla} \psi_{\rm CR} - \bfm{V}\,\psi_{\rm CR} \right)
\end{displaymath}
\begin{equation}
  ~~~~~~~~~ + \frac{\partial }{\partial p}\left[\frac{p}{3} (\bfm{\nabla}\cdot \bfm{V})\psi_{\rm CR} \right] - \frac{\psi_{\rm CR}}{\tau_{\rm CR}} + \mathcal{Q}_{\rm CR}\,,
\label{Eq_Diff_Adv}
\end{equation}
where $\psi_{\rm CR}({\bf r},p,t)$ denotes 
the CR density per unit of particle momentum $p$, at spatial position $r$.
Here $\mathcal{Q}_{\rm CR}$ is the cosmic-ray source term, $\mathcal{D}$ is the cosmic-ray 
diffusion coefficient, and $\tau_{\rm CR}$ is the cosmic-ray lifetime in the system.

For the advective flow of the gas, as motivated by observations, a divergence free 
velocity field is adopted, of the form

\begin{equation}
  \bfm{V} \cdot \hat{\textbf{z}} =v_{0} \times \frac{2\;{\rm e}^{\frac{1}{2}(1-\frac{d}{z})}}{1+z/d}\;,
\label{Eq_Breeze_Profile}
\end{equation}
where $v_{0}=300$\,km\,s$^{-1}$, $d=1$\,kpc, and $z$ is the distance to the Galactic plane. This profile broadly encapsulates 
the velocity profile of a \lq\lq breeze\rq\rq\/ solution for the isothermal outflow problem~\cite{Chamberlain:1965,Parker:1965}.

With regards the energy source driving this outflow, both a past AGN outburst event 
(see e.g.~\cite{Guo:2011eg,Guo:2011ip,Barkov:2013gda}), and a starburst 
phase or a sustained outflow driven by star formation in the Galactic centre 
(e.g.~\cite{Crocker:2014fla}) have been previously proposed. 
However, reference~\cite{Sarkar2016} claims that the present velocity 
data are not conclusive on the type of source responsible for this outflow. 
Energetically, the starburst-driven outflow luminosity is estimated to be
$\approx (1-3)\times 10^{40}$~erg~s$^{-1}$ \cite{Crocker:2014fla}. The present
level of AGN activity from the GC (of Sgr~A*) is considerably
below this ($L_{Sgr~A*}\sim 10^{33}$~erg~s$^{-1}$), but there is a growing body of evidence 
that its level in the recent past was significantly higher~\cite{Ponti:2010rn,Terrier:2010bn}. 
It therefore currently seems plausible
for either energy source to be driving the outflow.
We here choose to keep the discussion general, adopting instead the
velocity profile of Eq.~(\ref{Eq_Breeze_Profile}) as the starting point in our 
calculations.

We adopt a Monte Carlo approach to solve Eq.~(\ref{Eq_Diff_Adv}). Our results 
with this technique have been compared with those obtained using
an explicit differential equation solver. We found excellent agreement
in all cases.

We assume that the source term $\mathcal{Q}_{CR}$ is located at the GC region 
and constant in time. The copresence of the resultant accumulated CRs 
with the ambient gas gives rise to gamma-ray bubble 
emission through $\pi^{0}$ production generated in proton-proton interactions.
This emission may potentially account for the observed gamma-rays from 
the bubbles, as has previously been proposed by other authors~\cite{Crocker:2010}.

We determine the level of this emission by convolving the accumulated CR density 
throughout the outflow region with the target gas 
density in the outflow. As motivated on theoretical~\cite{Feldmann:2012rx}, 
and observational~\cite{Fang:2014hea} grounds, 
we adopt a constant gas density within the bubbles at the level 
$3\times 10^{-3}$\,cm$^{-3}$. In Fig.~\ref{gamma_contours}, we show a gamma-ray 
density map and a comparison of the 
gamma-ray bubble-edge profile with Fermi measurement. For these calculations, we took a CR luminosity 
of $10^{40}$\,erg/s for the central source. In Fig.~\ref{gamma_contours} (right panel), 
the origin of the diffuse gamma-ray emission in the $\theta>0$
region is assumed purely galactic in origin. Should some component of the
emission from this region be extragalactic, a reduction
of the GC luminosity or bubble gas density would be required in
order to account for such a reduction in required $\gamma$-ray emission intensity.

As can be seen in Fig.~\ref{gamma_contours}, a flat surface brightness
profile for the bubbles is obtained when assuming that
the velocity profile in the bubbles is described by Eq.~(\ref{Eq_Breeze_Profile}). 
We note, however, that in reality a range of velocity profiles 
can provide such a uniform brightness, see for example~\cite{Sarkar:2015xta}. 
In general, we find that for the case of a constant
density ambient gas description, the current gamma-ray data can 
be said to prefer decelerating profiles. For decreasing gas density 
profiles, a sharper fall-off in the velocity profile, than that 
used in Eq.~(\ref{Eq_Breeze_Profile}), would be required.

Although the cutoff at the edge of the bubbles is not well described by
the simple constant density gas model (see red dashed line in Fig.~\ref{gamma_contours}), 
a steeper cutoff in the $\gamma$-ray profile can be achieved by 
an abrupt change in the gas density at the 
bubble edge (see red solid line), as motivated in certain models~\cite{Crocker:2014fla}. 
Another motivation for such an origin for the bubble edges 
comes from a comparison of their morphology as seen 
in gamma-rays~\cite{Su:2010qj} and in radio~\cite{Carretti:2013sc}.
If GeV protons (resp. electrons) give rise to the $\gamma$-ray 
(resp. radio) emission, it would be surprising that the 
electrons extend out to larger latitudes than 
the protons. Such a difference between the morphologies of the $\gamma$-ray and radio data 
disfavours simple leptonic scenarios for the $\gamma$-ray bubbles. 
However, despite these challenges, more involved diffuse acceleration models 
supporting a scenario in which both the radio and gamma-ray emission are leptonic 
in origin are currently viable~\cite{Mertsch:2011es}.

\begin{figure}
\includegraphics[width=0.49\textwidth]{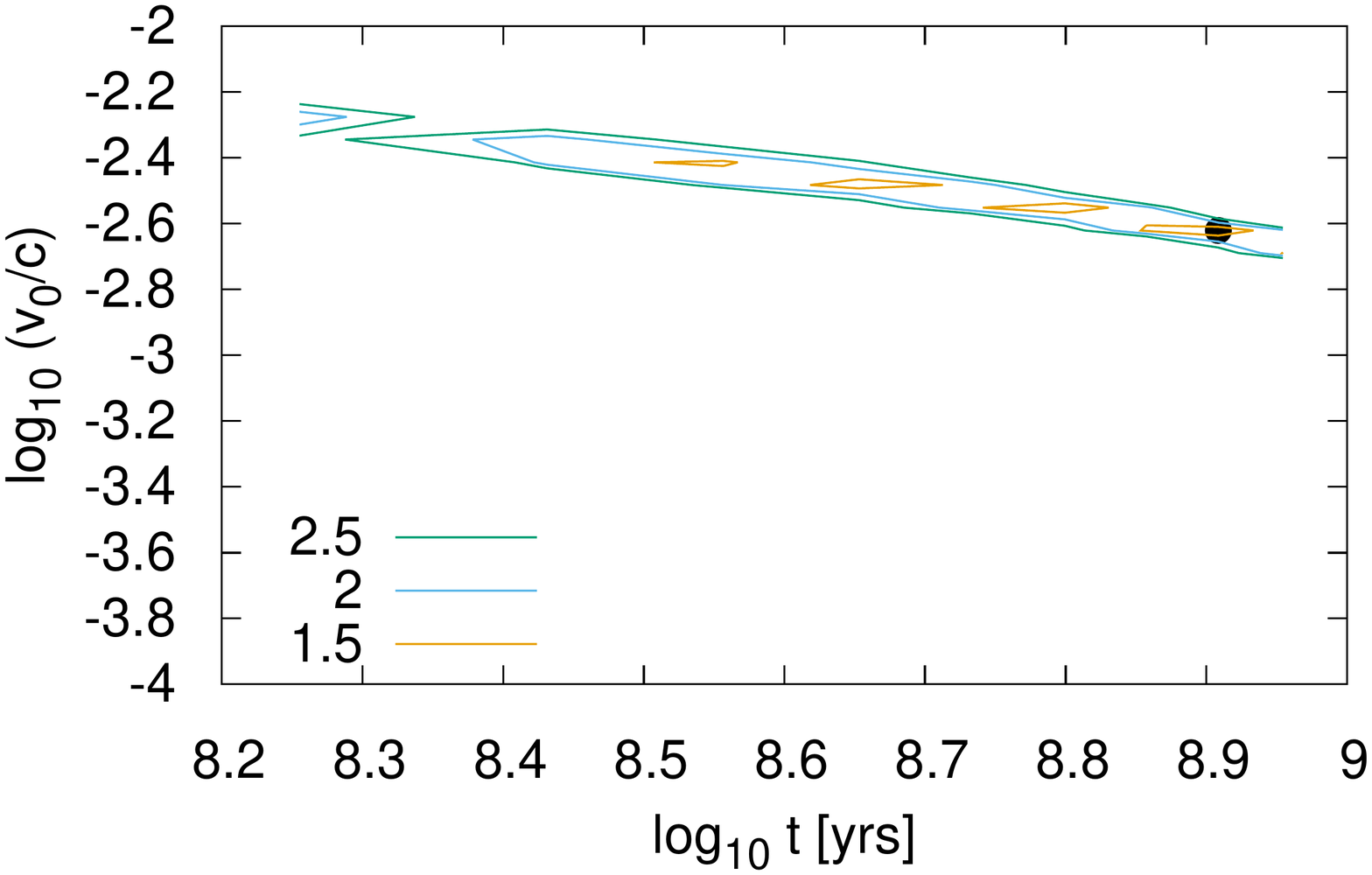}
\includegraphics[width=0.49\textwidth]{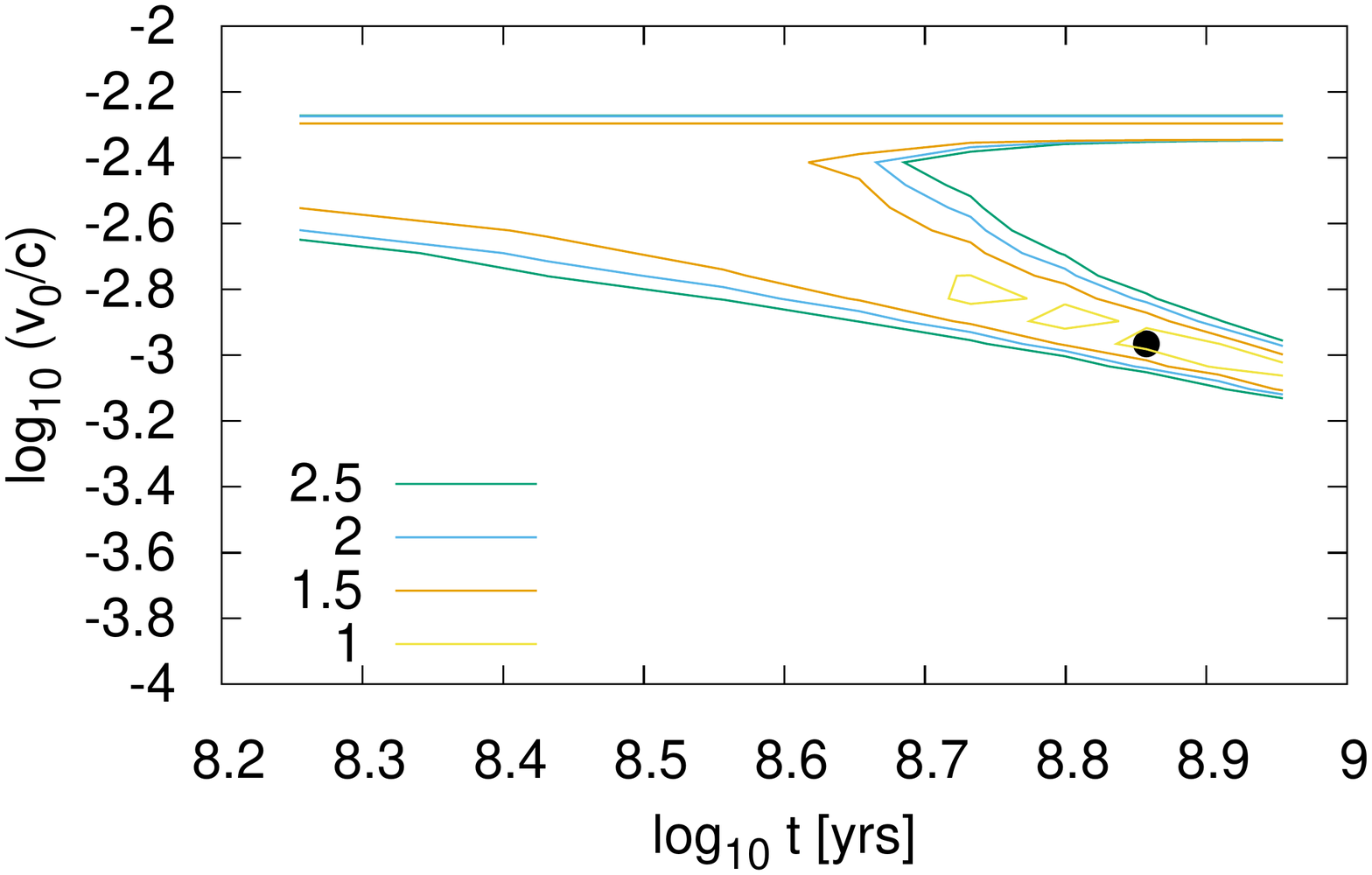}
\includegraphics[width=0.49\textwidth]{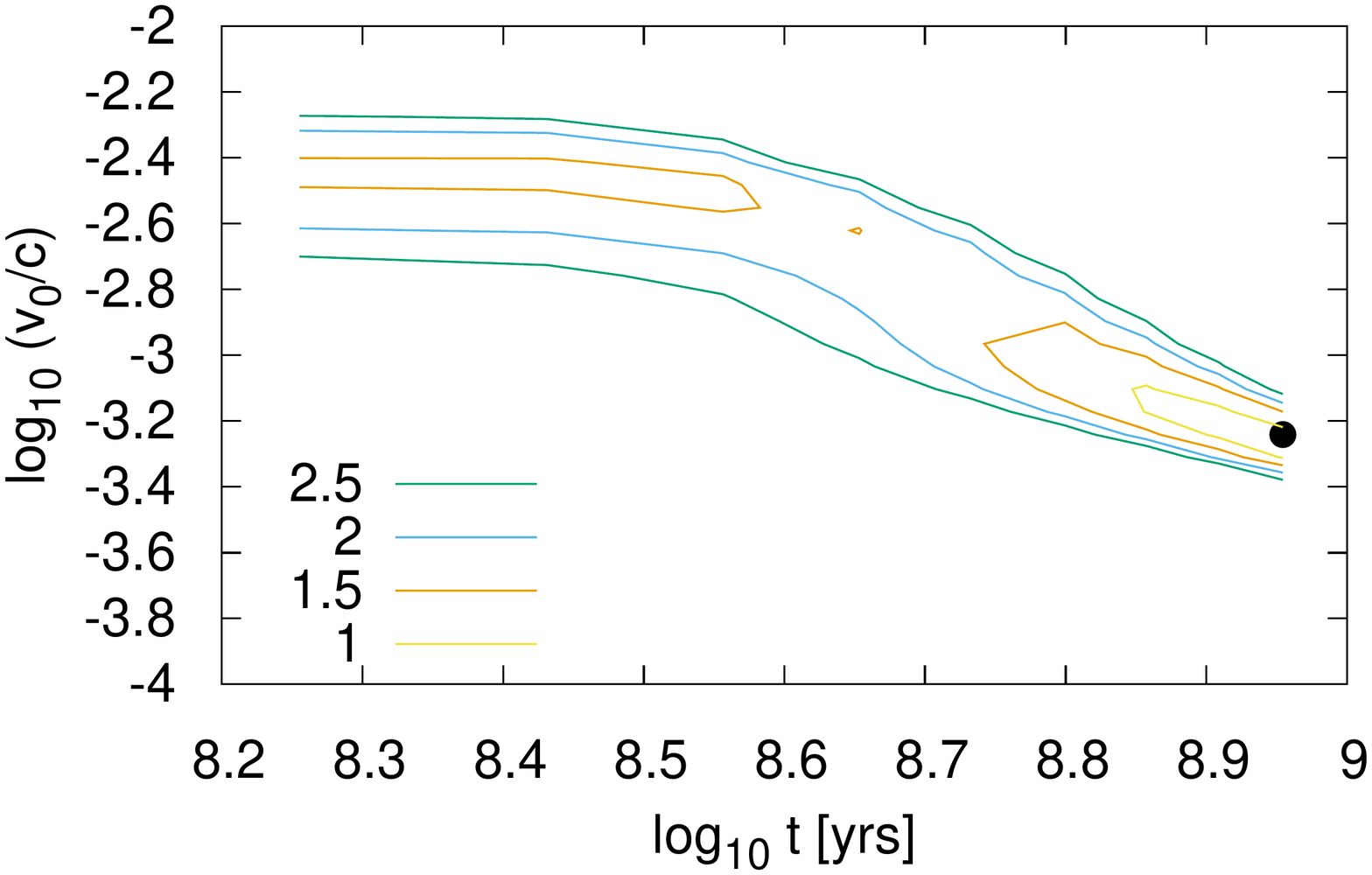}
\caption{Plots showing $\chi_{\rm d.o.f.}^{2}$ contours for fits to the $\gamma$-ray flux surface brightness profile of the Fermi-bubbles using Fermi satellite measurements in the range $\theta<2$[deg.]. The different contour plots cover a range of different cases for the distance $d$, over the range: $d=0.3$\,kpc (top); $d=1$\,kpc (middle); $d=2$\,kpc (bottom). In each plot, the position of the best-fit parameters is marked with a black circle.}
\label{Chi_contours}
\end{figure}

One simple explanation for the difference 
in the latitudinal profiles of the radio and gamma-ray emissions is that both 
protons and electrons possess extended distributions, and that the difference in morphology 
of their secondary emissions is due to differing distributions
of target gas and magnetic fields. A potential association of 
IceCube high energy neutrinos~\cite{Aartsen:2013jdh}, with the bubbles and 
beyond~\cite{Taylor:2014hya}, allows such a hadronic origin scenario for the 
gamma-rays to be tested in the near future.

With regards the parameter $d$, which dictates the turnover distance in the outflow
velocity profile described by Eq.~(\ref{Eq_Breeze_Profile}), a comparison of the 
fits to the radial gamma-ray profile of the Fermi bubbles is provided in
Fig.~\ref{Chi_contours}, through a consideration of the $\chi_{\rm d.o.f.}^{2}$ contours.
The upper plot in this figure shows that for the majority of the parameter
space, small values of $d$ are problematic, with the large $\chi_{\rm d.o.f.}^{2}$ values 
obtained reflecting the fact that such values lead to centrally brightened profiles, 
incompatible with the flat profile suggested by the data. 
However, the middle and lower panels show that the results for such 
intermediate and ``large'' values of $d$ both show considerable regions of
parameter space able to provide sufficiently flat profiles in agreement with
that measured.

\section{Local Outflow and CR Fluxes at Earth}
\label{CR_at_Earth}

In this Section, we study the impact that an outflow at larger Galactocentric radii would have on CR observables in the disk. The impacts of winds with either constant velocities~\cite{OwensJokipii1977,Jones1979,Genolini:2015cta} or velocities $V(z)$ increasing with height $z$ in the halo~\cite{Lerche1982,Dogiel1991,Bloemen1993,Recchia:2016ylf} have already been studied extensively in the literature. In the present work, we decide instead to study the case of outflows whose $dV/dz$ become negative above a given height $z_{\max}$. As far as we know, such a case has not been studied yet, with the exception of Reference~\cite{Taylor:2016qel}. We note that such velocity profiles do not correspond to those expected for winds driven by cosmic-rays. For such winds, $dV/dz \geq 0$ at all $z$, see Refs.~\cite{Ptuskin1997,Breitschwerdt:2002vs,Socrates:2006dv,Everett:2007dw,Samui:2009wk,Dorfi:2013cia,Zirakashvili2014,Recchia:2016ylf}, as well as the numerical simulations of Refs.~\cite{Hanasz:2013esa,Peters:2015yaa,Girichidis2016,Simpson2016}. However, the case of an outflow decelerating in the halo is worth studying for, at least, two reasons. First, some studies have argued that some galaxies may fail to produce \lq\lq successful\rq\rq\/, accelerating winds with a positive $dV/dz$ at all heights in the halo, for instance because of ram pressure from infalling material: See e.g. Reference~\cite{Dubois:2007wd}. Second, such a profile is preferred for the Fermi bubbles, as argued in the previous Section. It is unclear at the present time whether a breeze profile may apply to larger Galactocentric radii or not. Therefore, it is interesting to provide possible signatures that could confirm or rule out such a scenario.

As a first approximation, we assume in the following that variations of CR propagation or source parameters along the direction of the galactocentric radius can be ignored. We then assume that $V$ only depends on $z$. We wrote a code which solves numerically Equation~(\ref{Eq_Diff_Adv}) in planar 1D, for any arbitrary profiles of $V(z)$ and $\mathcal{D}(z,E)$. This code was presented and tested in~\cite{Taylor:2016qel}. In particular, we checked that it reproduces correctly the expected CR density profiles in the halo for the known cases of V=constant~\cite{OwensJokipii1977} and $V(z) \propto z$~\cite{Bloemen1993}, which are respectively constant and decreasing with $z$. On the contrary, the profiles $V(z)$ we consider below result in an increase of CR density above $z_{\max}$. Hereafter, we set $\psi_{\rm CR} = 0$ at $z=H$ as a boundary condition, where $H$ denotes the size of the halo. The CR density then decreases again when $z \rightarrow H$, due to CR escape. Such boundary conditions are widely used in Galactic CR propagation codes, and may correspond to the height above which the magnetic field is too weak to confine cosmic-rays. We stress however that, in general, the use of such conditions is not guaranteed to be justified, and may not be a good proxy for the actual physical picture: For instance, if a strong wind is present up to large $z$, see e.g. Refs.~\cite{Ptuskin1997,Zirakashvili2014}. In the latter case, the effective \lq\lq halo size\rq\rq\/ seen by GeV--PeV CRs corresponds instead to the (energy-dependent) height above which CR advection wins over CR diffusion. In the following, for our breeze profiles, $V(z)$ is small at large $z$, and we stick to the aforementioned boundary condition at $z=H$.

We calculate the steady-state distributions $\psi_{\rm CR}(z,E)$ for CR protons, and $\psi_{\rm B,C}(z,E)$ for boron and carbon nuclei. We calculate the production and destruction of boron as described in Ref.~\cite{Taylor:2016qel}, and we use the cross-section values quoted there. For the density profile of the target gas, $n(z)$, we take: $n=0.85$\,cm$^{-3}$ at $|z| \leq h$, and $10^{-3}$\,cm$^{-3}$ otherwise. The source term $\mathcal{Q}_{A}$ for primary nuclei $A$ is set to:
\begin{eqnarray}
\mathcal{Q}_{A} = \left\{ \begin{array}{ll}
                       \mathcal{F}_{A} \, \mathcal{Q}_{\rm CR}  & \mbox{, at } |z| \leq h~{\rm (disk)}\\
	               0  & \mbox{, at } h < |z| \leq H~{\rm (halo)}
                       \end{array} \right.
\label{Q_A}
\end{eqnarray}
where $\mathcal{F}_{\rm A}$ denotes the fraction of nuclei $A$ emitted at the sources. The disk width is set to $h=200$\,pc. For clarity, we assume below that there are no sources of primary boron. See e.g.~\cite{Genolini:2015cta} for a non-zero $\mathcal{Q}_{\rm B}$. We will not study here the hypothetical case where CR trapping around their sources contributes significantly to the boron-to-carbon ratio, e.g.~\cite{D'Angelo:2015clt}. In such a scenario, this ratio would contain little information on CR propagation on large scales in our Galaxy, and hence little constraints on a local outflow. For clarity, we assume below that $\mathcal{D}$ does not depend on $z$. We express it as:
\begin{eqnarray}
\mathcal{D}(E/Z) = \mathcal{D}_{3\,{\rm GV}}\, \left( \frac{E/Z}{3\,{\rm GV}} \right)^{\delta}\;,
\label{D}
\end{eqnarray}
for nuclei of charge $Z$. We set $\delta=0.44$ and $\mathcal{D}_{3\,{\rm GV}} = 2.8 \times 10^{28}\,{\rm cm}^2\,{\rm s}^{-1}$, which correspond to the best fit values of Ref.~\cite{Genolini:2015cta} for $H=4$\,kpc. Ref.~\cite{Evoli:2013lma} also suggested the same value for $\delta$. We verified that our code reproduces the expected boron-to-carbon ratios both for the cases of \lq\lq no wind\rq\rq , and \lq\lq wind velocity constant with $z$\rq\rq .

In the static regime ($V=0$), the boron current in the halo is $\mathcal{J}_{\rm B,Halo} = -\mathcal{D}_{\rm B} \partial \psi_{\rm B}/\partial z \simeq \mathcal{D}_{\rm B}\psi_{\rm B,0}/H$ ($h \ll H$). Therefore, in the static regime, the boron-to-carbon ratio is
\begin{eqnarray}
  \frac{\psi_{\rm B,0}}{\psi_{\rm C,0}} \simeq \frac{\sigma_{\rm \rightarrow B}}{\sigma_{\rm B \rightarrow} + \frac{\mathcal{D}_{\rm B}(E)}{cn_{0}hH}} = \frac{\tau_{\rm \rightarrow B,0}^{-1}}{\tau_{\rm B \rightarrow ,0}^{-1} + \frac{\mathcal{D}_{\rm B}(E)}{hH}}\;,
\label{BoC_1}
\end{eqnarray}
where indices \lq\lq 0\rq\rq\/ mean \lq\lq at $z=0$\rq\rq, and $\sigma_{\rm \rightarrow B,\,B \rightarrow}=1/cn\tau_{\rm \rightarrow B,\,B \rightarrow}$ are the production and destruction cross-sections for boron. With the parameter values we take here, the diffusion term \lq\lq $\frac{\mathcal{D}_{\rm B}(E)}{hH}$\rq\rq\/ dominates over the \lq\lq $\tau$\rq\rq\/ term only around the last couple of points in the AMS-02 data~\cite{AMS2013}, which is why the slope in the data, at $\lesssim {\rm a~few} \times 100$\,GeV/nucl, looks flatter than 0.44. This calls for a better knowledge of cross-sections, as also noted by~\cite{Genolini:2015cta}.

In the hypothetical case of a wind with a velocity constant with $z$, the \lq\lq $\frac{\mathcal{D}_{\rm B}(E)}{hH}$\rq\rq\/ term in Eq.~(\ref{BoC_1}) must be replaced with \lq\lq $V/(1- \exp [-HV/\mathcal{D}_{\rm B}(E)] )h$\rq\rq, cf. Ref.~\cite{Genolini:2015cta}. Let us denote $z_{\ast}=\mathcal{D}/V$, the distance beyond which advection dominates over diffusion. In this particular setup, low-energy CRs with $z_{\ast}=\mathcal{D}/V<H$ advect to the boundary, whereas higher-energy CRs diffuse to the boundary. This introduces a flattening in the boron-to-carbon ratio at low energies ($z_{\ast} \propto E^{\delta}$ for $V=$~constant). A value of $V$ larger than a few tens of km/s for $H\sim 10$\,kpc is incompatible with the data. This excludes a strong \lq\lq $V=$~constant\rq\rq\/ wind. However, the current boron-to-carbon data does {\em not} exclude the presence of a strong wind in general, as other wind profiles with $V\neq$~constant, such as $V(z)\propto z$, are allowed by the data. For winds with $V(z)\propto z$, $z_{\ast}\propto \sqrt{\mathcal{D}}$. For a CR spectrum at the sources $\propto E^{-\alpha}$, the slope of the CR flux at Earth then tends to $-\alpha-\delta$/2, cf. Reference~\cite{Bloemen1993}.

\begin{figure}
\includegraphics[width=0.49\textwidth]{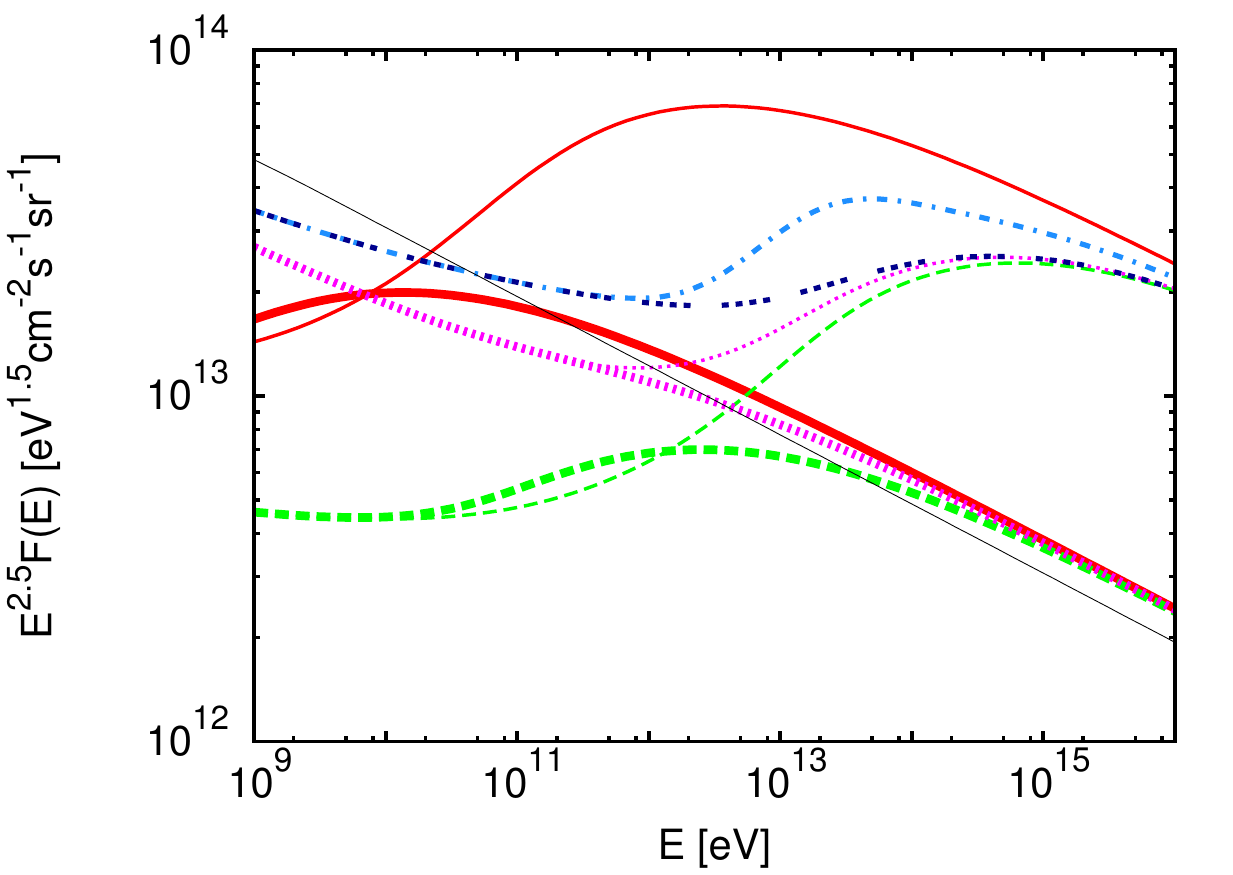}
\includegraphics[width=0.49\textwidth]{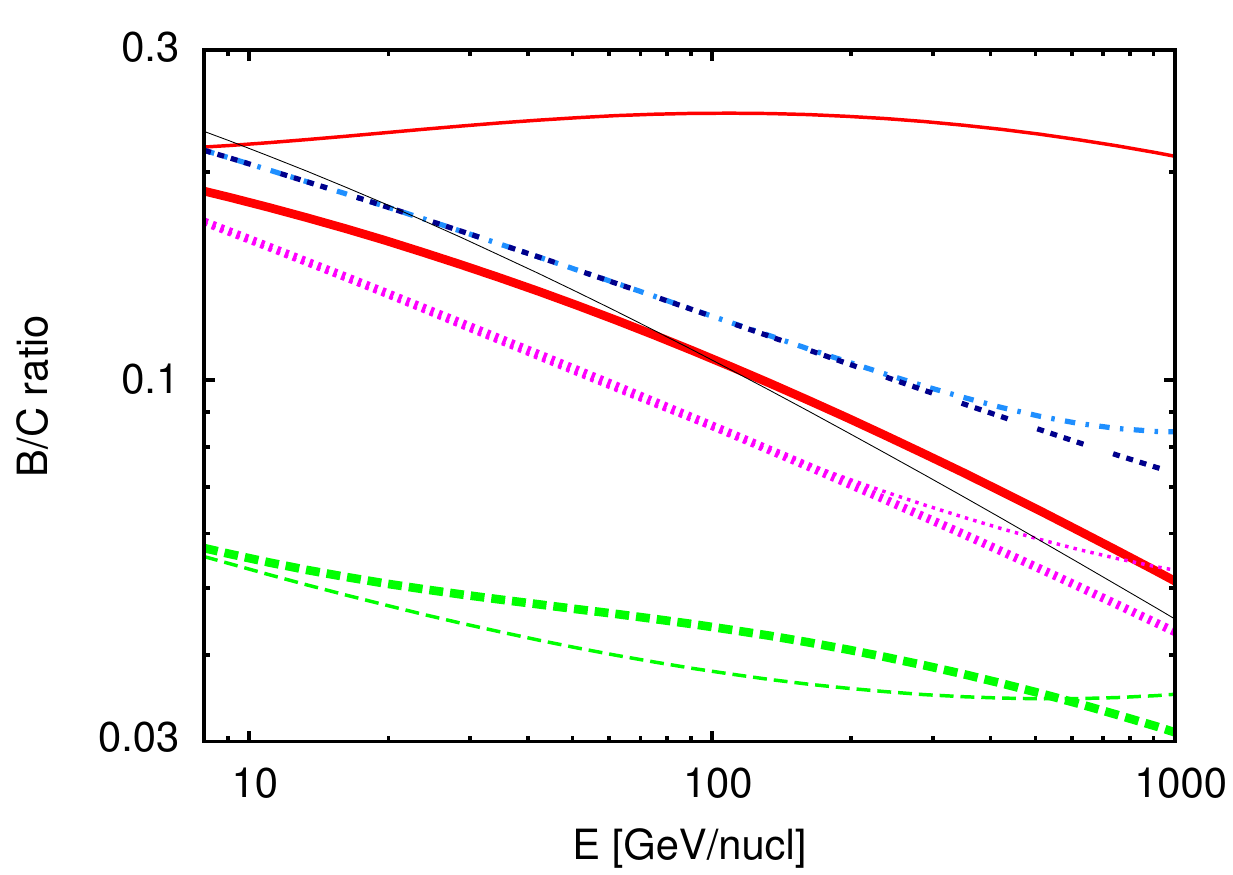}
\includegraphics[width=0.49\textwidth]{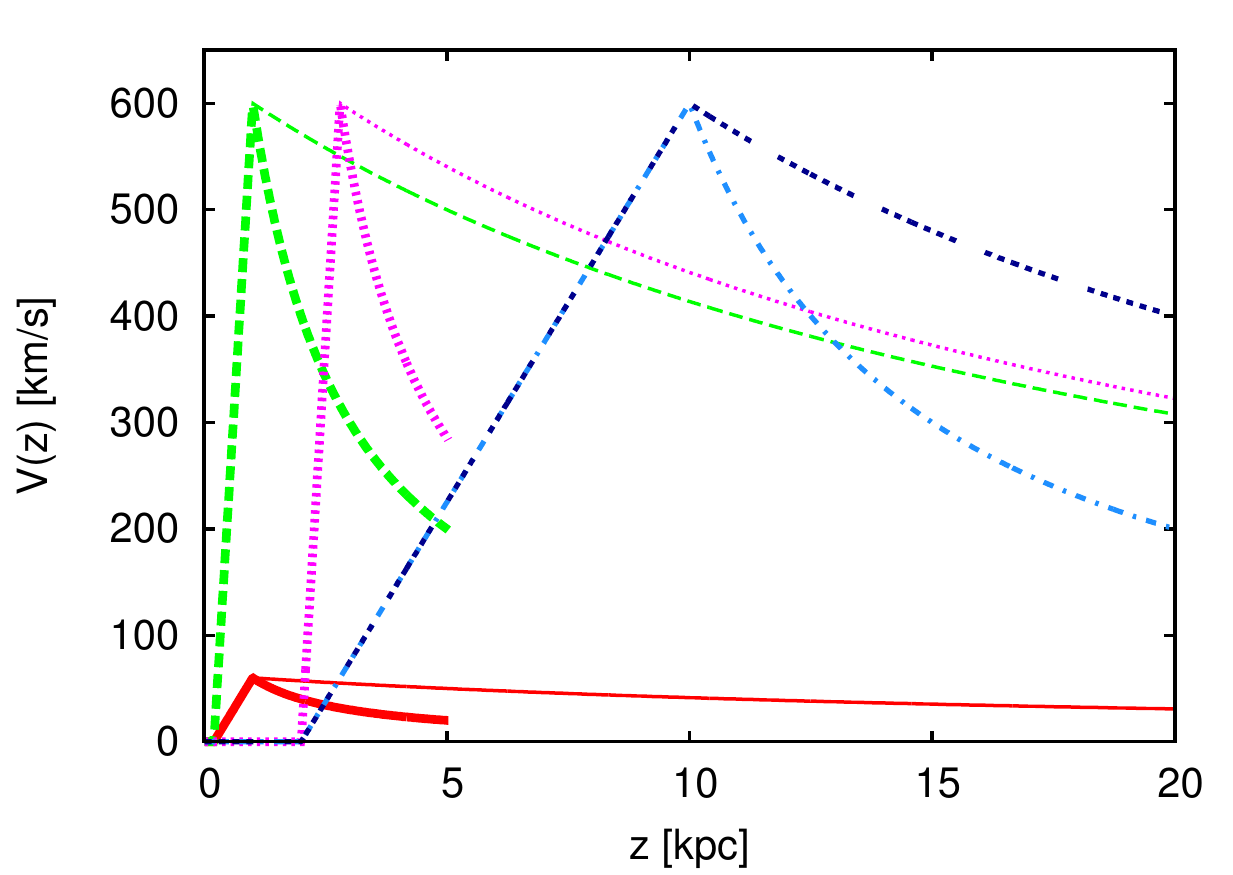}
\caption{CR flux ({\it upper panel}) and B/C ratio ({\it middle panel}) at $z=0$, for the outflow profiles displayed in the {\it lower panel} ($20\,{\rm kpc} < z \leq 50\,{\rm kpc}$ not shown). We do {\it not} try to fit the data. $\mathcal{D}_{3\,{\rm GV}} = 2.8 \times 10^{28}\,{\rm cm}^2\,{\rm s}^{-1}$, $\delta = 0.44$, $h=200$\,pc, $H=5$ or 50\,kpc (see end of lines on the right panel), $n(z)=0.85$\,cm$^{-3}$ for $|z| \leq h$ and $10^{-3}$\,cm$^{-3}$ otherwise, CR spectrum at sources $\propto E^{-2.26}$, and total power injected in CRs at $|z| \leq h$ in this region of the disk set to $\approx 3.3 \times 10^{39}$\,erg\,pc$^{-2}$\,yr$^{-1}$. Each scenario is represented by the same line type on each panel, see text. Thin black line for $V=0$, $H=4$\,kpc, and other parameters unchanged.}
\label{Profiles_1}
\end{figure}

From now on, we focus on $V(z)$ profiles that decrease above a height $z_{\max}$ in the halo. In Figure~\ref{Profiles_1}, we show eight profiles (lower panel) and their impact on the CR spectrum (upper panel) and boron-to-carbon ratio (middle panel) at $z=0$. For reference, we also plot results for the \lq\lq benchmark fit\rq\rq\/ of~\cite{Genolini:2015cta} for $V=0$ (thin black line). The boron-to-carbon ratio measurements from AMS-02 experiment coincide with this line. Our goal here is {\em not} to provide a fit of the data. Instead, we take \lq\lq caricatural\rq\rq\/ examples of $V(z)$ profiles with rather extreme parameter values, so as to make the impact of these parameters more visible. In some cases, changing the values of some of the parameters would provide a reasonable fit to the data.

We set the CR spectral index at the sources to $\alpha=2.26$. We do not vary $\alpha$, $\delta$, $\mathcal{D}_{3\,{\rm GV}}$, $n(z)$ or $h$, so as to help the reader distinguish between the contributions from the different wind parameters. For all eight profiles, $V=0$ at $z\leq z_{\min}$, then increases $\propto z$ up to $V_{\max}$ at $z=z_{\max}$, and then decreases on a typical length scale $d'$:
\begin{eqnarray}
V(z) = \left\{ \begin{array}{ll}
                       0 &\mbox{, at } |z| \leq z_{\min}\\
	               \frac{V_{\max}(z-z_{\min})}{z_{\max}-z_{\min}} &\mbox{, } z_{\min} < |z| \leq z_{\max}\\
	               \frac{V_{\max}}{1+(z-z_{\max})/d'} &\mbox{, } z_{\max} < |z| \leq H
                       \end{array} \right.
\label{V_z_Profiles}
\end{eqnarray}

Each of these scenarios is represented by the same line type and colour in each panel of Fig.~\ref{Profiles_1}. Let us denote by A (resp. B) the thick (resp. thin) red solid lines in Fig.~\ref{Profiles_1}, C (resp. D) the thick (resp. thin) green dashed lines, E (resp. F) the thick (resp. thin) magenta dotted lines, and G (resp. H) the dark (resp. light) blue dashed-dotted lines. We take $z_{\min}=200$\,pc for \{A,B,C,D\}, and 2\,kpc otherwise. $z_{\max}=1$\,kpc for \{A,B,C,D\}, 2.8\,kpc for \{E,F\}, and 10\,kpc for \{G,H\}. $V_{\max}=60$\,km/s for \{A,B\}, and 600\,km/s otherwise. $d'=2$\,kpc for \{A,C,E\}, 5\,kpc for H, and 20\,kpc otherwise. $H=5$\,kpc for \{A,C,E\}, and 50\,kpc otherwise.

In all cases, when the CR energy is sufficiently high for diffusion to win over advection at all $z$, the problem simplifies to a basic leaky-box with homogeneous diffusion coefficient. Spectra (upper panel) then all tend to $\propto E^{-\alpha - \delta = -2.7}$, e.g. above $\sim 100$\,GeV (resp. $\sim 10$\,PeV) for A (resp. G). Case A shows that a weak outflow with such a profile introduces a turnover in the CR spectrum at low energies. This might explain the turnover measured below $\sim 10$\,GeV from molecular clouds. The outflow reduces the \lq\lq box\rq\rq\/ (in which low-energy CRs diffuse and can come back to $z=0$) to an effective size $z_{\ast} \ll H$, where $z_{\ast}(E)$ is the height where advection wins over diffusion. For these low-energy CRs, advection dominates in the region of large $V$ around $z=z_{\max}$. This region then acts as a bottleneck for them. A small turnover also occurs in the boron-to-carbon ratio. When increasing the extent of the wind, and then the size $\Delta z$ of the region in the halo in which $V$ is greater than a fraction of $V_{\max}$, the turnover is shifted to higher energy: Indeed, in this case, a larger $\mathcal{D}(E)$ is required for diffusion to win over advection, see curves B. The CR flux at high energies in case B is ten times larger than in case A, because we kept $\mathcal{D}_{3\,{\rm GV}}$ fixed in these examples, and did not rescale it with $H$.

Increasing $V_{\max}$ also increases the energy of the aforementioned feature, see C and D. More interestingly, CRs below this energy that can come back to $z=0$, probe the part $\propto z$ of the wind, because $z_{\ast} \leq z_{\max}$ for them. Since $z_{\max} \gg z_{\min} (\approx 0)$ in C and D, this results in a \lq\lq $V(z) \propto z$\rq\rq\/ scenario~\cite{Bloemen1993}, and the CR spectrum (resp. boron-to-carbon ratio) slope tends to $-\alpha-\delta$/2 (resp. $-\delta$/2) at low energy, see green lines in Fig.~\ref{Profiles_1}. On the contrary, if one increases the altitude $z_{\min}$ of the launching of the wind, the CR spectrum (resp. boron-to-carbon ratio) slope tends to $-\alpha-\delta$ (resp. $-\delta$), as for the \lq\lq $V=0$\rq\rq\/ scenario: See the magenta lines for the cases E and F. This is due to the fact that, at these low energies, $z_{\min} \gg z_{\ast} - z_{\min}$. Therefore, low-energy CRs see a small, energy-independent, effective leaky-box of height $\approx z_{\min}$. In the cases E and F, $z_{\max}-z_{\min} \ll z_{\min}$, and, with increasing CR energy, there is a \lq\lq quick\rq\rq\/ transition to a bigger box of effective size $\approx H$: The slopes of the magenta CR spectra are $\approx -\alpha-\delta$ both at low and high energies in the energy range displayed in the upper panel.

For the case G, there is a hint of a smooth transition, at low energy, from a $\sim -\alpha-\delta$ to a $\approx -\alpha-\delta$/2 slope. Indeed, $z_{\min}$ has the same large value as in case F, but, for G, $z_{\max}-z_{\min}$ is not small compared to $z_{\min}$. It is interesting to note that G and H are not far from fitting the existing experimental data for the boron-to-carbon ratio, despite having velocity profiles very different from the $V=0$ profile of the \lq\lq benchmark fit\rq\rq\/ (thin black line). The impact of $d'$ on the CR spectrum is visible by comparing G with H. H, the profile with a faster fall off at high $z$, achieves the transition between the two limiting regimes in a smaller energy interval. A hardening (upturn) is also present in the boron-to-carbon ratio for the cases D, F, and H.

Green, magenta, and blue lines show a hardening in the CR spectrum around $\sim 10^{10-13}$\,eV, due to the launching of a wind in the halo. For some parameter values, it is possible to make it coincide better with the one measured at 200\,GV in the CR spectra by PAMELA, CREAM and AMS-02 experiments. Thence, even in the limiting case of equal $\mathcal{D}$ in the disk and in the halo, such a hardening may arise from the launching of a breeze or wind. This argument is valid for cases with $dV/dz > 0$ above $z_{\max}$ too. As noted above, similar hardenings also appear in the boron-to-carbon ratio. This does not contradict the experimental data, provided that the hardening is concealed at higher energies, or remains within the systematics of the detectors. Regarding the second of these possibilities, conflicts in secondary to primary ratios have been reported in existing data sets, cf. Ti/Fe ratio by HEAO-3-C3~\cite{Vylet:1990}, ATIC-2~\cite{Zatsepin:2009}, and comparison to the boron-to-carbon ratio~\cite{AMS2013}.


\section{Conclusions and perspectives}
\label{conclusion}


We presented in Section~\ref{GC_outflow} a hadronic model of the Fermi bubbles. Assuming that they result from a Galactocentric outflow carrying pre-accelerated cosmic-rays, we calculated the gamma-ray emission produced by the CRs interacting with the gas present in the bubbles. We showed that outflows decelerating with distance to the Galactic disk can reproduce the flat gamma-ray surface brightness of the bubbles, in accordance with the measurements from Fermi satellite. Our description for the outflow profile is enclosed in breeze solutions of isothermal winds.

Motivated by the above findings, we studied in Section~\ref{CR_at_Earth} the impacts that similar types of outflow profiles would have on the CR spectra at Earth, should such outflows exist at larger Galactocentric radii. Competition between CR diffusion and advection in the halo can produce an inflection point in the CR spectrum at $z=0$. A hardening can appear in the CR spectrum due to the launching of a wind or breeze in the halo, even in the hypothetical, limiting case of equal CR diffusion coefficients in the halo and disk.

Although a breeze outflow scenario is currently only motivated for the outflow from the GC region, we conclude from the above results that future observations should be able to test its presence or absence at larger radii, thanks to local CR observables.

\section*{Acknowledgments}
AT acknowledges a Schroedinger fellowship at DIAS.

\nocite{*}
\bibliographystyle{elsarticle-num}
\bibliography{references1}




\end{document}